\documentclass[twocolumn,twoside,slac]{revtex4}
\usepackage{graphicx}
\usepackage{fancyhdr}
\def\lbabar{\mbox{\sl {\Large\sl B}\hspace{-0.45em} A\hspace{-0.1em}{\Large\sl B}\hspace{-0.45em} A\hspace{-0.1em}R}}
\def\babar{\mbox{\sl B\hspace{-0.4em} {\scriptsize\sl A}\hspace{-0.4em} B\hspace{-0.4em} {\scriptsize\sl A\hspace{-0.1em}R}}}

\begin{document}

\title{Distributed Offline Data Reconstruction in \lbabar}

%

\author{Anders Ryd}
\affiliation{California Institute of Technology, Pasadena, CA 91125}
\author{Alberto Crescente, Alvise Dorigo, Fulvio Galeazzi, Mauro Morandin, Roberto Stroili, Gionni Tiozzo, Gabriele Vedovato}
\affiliation{INFN Padova, I-35131 Padova, Italy}
\author{Francesco Safai Tehrani}
\affiliation{INFN Rome, I-00185 Rome, Italy}
\author{Teela Pulliam}
\affiliation{Ohio State University, Columbus, Ohio 43210}
\author{Peter Elmer}
\affiliation{Princeton University, Princeton, NJ 08544}
\author{Antonio Ceseracciu, Martino Piemontese}
\affiliation{SLAC, Stanford 94025 /INFN Padova, I-35131 Padova, Italy}
\author{Doug Johnson}
\affiliation{University of Colorado, Boulder, CO 80309}
\author{Sridhara Dasu}
\affiliation{University of Wisconsin, Madison, WI 53706}
\author{ }
\affiliation{ }

\author{For the \babar\ Computing Group}

\begin{abstract}
The \babar\ experiment at SLAC is in its fourth year of running. The data 
processing system has been continuously evolving to meet the challenges of 
higher luminosity running and the increasing bulk of data to re-process each 
year. To meet these goals a two-pass processing architecture has been adopted, 
where 'rolling calibrations' are quickly calculated on a small fraction of the 
events in the first pass and the bulk data reconstruction done in the second. 
This allows for quick detector feedback in the first pass and allows for the 
parallelization of the second pass over two or more separate farms. This two-pass 
system allows also for distribution of processing farms off-site. The first such 
site has been setup at INFN Padova. The challenges met here were many. The software 
was ported to a full Linux-based, commodity hardware system. The raw dataset, 90 $TB$, 
was imported from SLAC utilizing a 155 $Mbps$ network link. A system for quality control 
and export of the processed data back to SLAC was developed. Between SLAC and Padova 
we are currently running 
three pass-one farms, with 32 CPUs each, and nine pass-two farms with 64 to 80 CPUs 
each. The pass-two farms can process between 2 and 4 million events per day. Details 
about the implementation and performance of the system will be presented.
\end{abstract}

\maketitle

\thispagestyle{fancy}


\section{The \lbabar\ Experiment}
\babar\ is an experiment built primarily to study B-physics at an 
asymmetric high luminosity electron positron collider (PEP-II) at the 
Stanford Linear Accelerator Center (SLAC).
It is an international collaboration involving 560 physicists from
76 institutions in 10 countries. 

\babar\ has been taking data since May 1999 and is currently in 
the middle of ``Run3'', which will run through June 2003.
To date it has collected about $110 fb^{-1}$ of data which 
corresponds to about 1.1 billion fully reconstructed events. 
It is expected that the
data sample will increase to the order of $500 fb^{-1}$ by the
end of 2006.

\section{Overview of the data reconstruction process}
\begin{figure}
\vspace{-4mm}
\centering
\includegraphics[width=70mm]{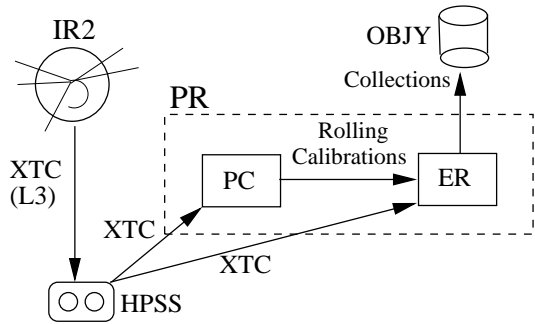}
\caption{Overview of Data Processing. The raw data collected by the
detector at IR2 is written to XTC files which are stored in the HPSS
tape system. The XTC files are then processed twice, once to produce
the calibrations (PC) and once to do the full data reconstruction (ER),
where the reconstructed events are written to the event-store database
(Objy).} \label{data_path}
\end{figure}

A diagram outlining the data storage and processing system
is shown in figure~\ref{data_path}.

\subsection{Raw Data}
The raw data coming from the \babar\ detector (located at Interaction 
Region 2 (IR2)) and filtered through a L3 trigger
are written to flat files, called XTC (eXtended Tagged Container) files which are 
stored in a mass storage (HPSS) system shortly after production. 
The size of the raw data per event is about $30kB$. The average size for 
an XTC file is about $10GB$, containing about 300,000 events. In the
past we have written files containing up to $1M$ events.

Each XTC file contains all the events taken for a single run 
of the collider and all the events of a single XTC file are processed together. 
We record more data than we fully process. All but 35-40\%\ of the events
in the XTC file are rejected by dedicated filters early in the 
reconstruction executable before full reconstruction is performed. 

\subsection{Data Processing Overview}
The data is processed in a two pass Prompt Reconstruction (PR) system, 
described in detail in Section~\ref{TwoPass}.
The XTC files are read twice, once to calculate the calibrations and then to fully
reconstruct the data. The fully reconstructed physical quantities are written into 
an object database (Objectivity/DB ~\cite{objy-ref}). The processed data is then transfered to 
a separate database where users can access it~\cite{artem-ref}.

\subsection{Reconstructed Data}
The output of the reconstruction is written to an event-store 
database. The following quantities can be written, per event:
tag, micro, mini, reco and raw.

The tag and micro quantities contain the highest level information
about the reconstructed event. The tag information records simple 
event selection criteria, such as number of charged tracks in the event. 
The micro includes standard information about the event and reconstructed
tracks and composites. These two quantities have been the primary format
used for analysis in \babar.

The mini and reco level contain more detailed information
about the reconstructed candidates in the event in order to allow
users to redo some of the reconstruction if needed.
At the end of 2001 we deprecated writing the reco level information
($100kB$+/event) to the event-store. It's function will be largely replaced
by a redesigned mini~\cite{mini-ref}.

In summer 2002 we also deprecated writing a copy of the raw data to
the event-store ($50kB$/event). The original purpose of writing the raw
data was to enable reprocessing directly from the event-store
database instead of from the XTC files, but this was never pursued. 

The current output size per event in the event-store (tag/micro/mini)
is about $20kB$. Physics selections are run as part of the reconstruction
and currently 4 physical streams are written to the 
event-store along with 111 pointer skims. This pre-selects events in 
categories useful for physics analysis and enables the user to easily run 
only on a specific subset of the data.

The Objectivity/DB database that stores the reconstructed data can contain
runs processed more than once with different software versions or different
calibrations or just because something went wrong the previous time.

The reconstructed events for each run are organized inside the database 
in collections. There are 115 collections per run, one for each output skim or stream.
The collection name is constructed to contain all the unique information about
the collection; the stream name, the software release version used to process
the data, and the run number. This is an example:

\noindent/groups/AllEvents/0001/3000/P12.3.4aV06fb
/00013026/cb001/allevents.

``AllEvents'' specifies one of the physical streams which contains all events which 
pass a loose physics selection. P12.3.4aV06 specifies the release used (``12.3.4a''), 
that it was a Production release (leading ``P''), and that the run has been processed 
7 times (``V06''). The run number is for this run is 13026.

\subsection{Reprocessing}
As in any active experiment the data reconstruction algorithms and the 
detector calibrations are continuously being improved as our understanding
of the detector increases. In order for the physics measurements to benefit 
from these improvements it is necessity to reprocess the accumulated data set
each year from the raw (XTC) files.

The total throughput needed for reprocessing may actually exceed that needed
for processing new data. The capacity needed is defined by the time a
stable reconstruction executable becomes available and,the deadline by which 
data must be reprocessed (e.g. for analysis in time for specific conferences),
and the current data sample.

Scaling for reprocessing can be accomplished by breaking the 
conditions time-line into separate intervals and creating a 
separate instance of the two-pass PR system for each time
interval. The calibrations are then calculated within each
separate interval and the separate run ranges can be processed 
in parallel.

A sophisticated book-keeping system, based on a relational SQL database 
(Oracle~\cite{oracle-ref} or MySql~\cite{mySql-ref}), 
keeps track of all processing and reprocessing jobs. It records the date, time,
software release, and calibration used for that (re)processing of the data, as
well as other statistical quantities.

\section{Distributed Processing}
Given the large number of events per run (XTC file) it is 
not practical to process the entire run on a single CPU,
thus a parallel processing architecture has been implemented.
As described in detail below, a central server reads the 
events from the XTC file and distributes them to a set of 
client processes. 

\begin{figure}
\vspace{-4mm}
\centering
\includegraphics[width=70mm]{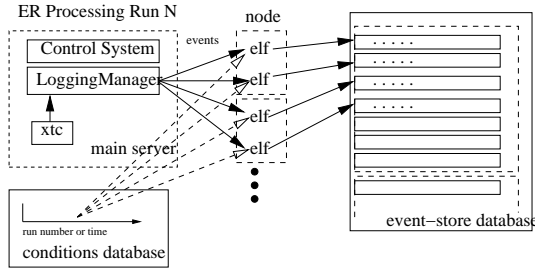}
\caption{Sketch of an ER Processing Farm. The Control System (CS) and Logging Manager (LM)
run on the main server. The LM reads the XTC file from a local disk and send out events to
the reconstruction code (elf) running on the farm nodes. The elves read the conditions
from the conditions database and write out reconstructed events to the event-store database.} 
\label{farm}
\end{figure}

One processing farm consists of a main server, a number of farm nodes,
and a number of Objectivity/DB servers. The main server has a large local (SCSI)
disk ($200-250GB$) where the XTC files are staged and logfiles are written to
with a gigabit network connection.
The nodes are currently dual-processor linux boxes
with a fast-ethernet network connection.
The Objectivity/DB servers are Solaris machines at SLAC, and Linux 
machines in Padova, which include a 
lockserver, journal server and a datamover with 1TB raid
arrays to store the processed events or calibrations.
Figure~\ref{farm} gives a rough picture of an ER processing farm.

\subsection{The Logging Manager}
Special software developed by \babar\ called the Logging Manager (LM)~\ref{lm-ref} runs on a dedicated 
machine (the main server), reads the events sequentially from a single XTC file and then 
distributes them in parallel to many hosts (dual processor nodes). On a single
node two instances of the reconstruction code (Elf) request events from the LM
through TCP/IP, 
process them and write the output to the event-store database. 

The LM keeps track of the events sent to each Elf
and if an Elf crashes without processing them or committing them to the event-store
it can resubmit these events to another Elf.
The LM makes sure that each event in the XTC file is processed and written out once and only
once. If an Elf crashes while processing a particular event, that event is
tagged, by the LM, as a ``killer event'' and not redistributed to the other
elves. These ``killer events'' are usually due to problems in the reconstruction
code which can not handle events with certain components. These problems are
monitored for and fixed with new releases of the reconstruction code. 

\subsection{The core of the reconstruction: Elf}
The reconstruction code (Elf) runs locally on the nodes.
While processing the first event assigned to it by the LM, each 
Elf reads calibrations from the conditions database. As
the events are processed, it stores them in a memory buffer on the node; 
when the buffer is full Elf flushes everything into the 
event-store database and then requests new events from the LM.. 

The number of events stored in the 
buffer before writing into the event-store is tuned by two parameters: 
a memory cache size and a commit time interval. Since all the nodes
in a farm (60-80 nodes) all write to the same event-store database the
commit time is made as random as possible, to avoid collisions between
multiple nodes trying to write to the same area in the database.

\subsection{The Event-Store and the Clustering Hint Server}
The Objectivity/DB event-store database into which Elf writes the events
consists of database files managed by a general catalog. The databases, 
the catalog and other metadata files, are hosted by dedicated 
machines (called ``datamovers'') and are accessible via an object oriented engine provided
by Objectivity/DB in collaboration with SLAC: the AMS server. Different database files correspond to different 
categories of data: raw data, particle tracks, kind of physics, skims, metadata, 
collections, etc. Internally, a single database file is organized by containers 
(the smallest lockable unit inside the database).
 
Before Elf flushes its cache into the event-store, it asks for containers to write into 
(which can belong to different database files) and locks them (using a special 
lock server provided by Objectivity/DB). This operation is done by many clients and 
is fairly frequent and can easily constitute a bottleneck, unless the
commit time interval is properly chosen as explained in the previous paragraph.

To help with this problem SLAC software engineers wrote a Clustering 
Hint Server (CHS) on top of the Objectivity/DB system. The CHS is a deamon running on 
a dedicated machine which continuously communicates with the AMS server and scans the 
event-store federation. It keeps in memory all the container addresses, analogous to a 
huge hash table. Before Elf writes events into the event-store 
it asks the CHS for a free container which the CHS provides very quickly.
The CHS also pre-creates database files in the background when the existing ones are almost full 
(each database file has a maximum size fixed to $1.8GB$), and creates their internal containers; 
so Elf always has free containers available to write into.

\section{The PR control system}
All low-level components described above (e.g. LM, Elf) are controlled by a high level 
Control System (CS). Its task is to start/stop all the low-level executables, submit runs 
for reconstruction, update the bookkeeping database, 
report to the shifters the processing status in run-time,
send emails and pages to experts and shifters in case of errors or anomalies, 
and produce histograms of some reconstructed detector and physics quantities 
for data quality assurance.

The original version of the CS was developed incrementally 
and used very successfully during the first three years of data taking. 
The old CS architecture was client-server 
based, see ~\cite{franz-ref} for a detailed discussion of the architecture.

For the fourth year of data taking it was decided to redesign the CS, building 
upon the experience of the first three years.
The old control system was developed during production where solutions to
new problems needed to be fixed in the quickest way possible, which is not usually
the best or cleanest way. (See ~\cite{tom-ref} for a discussion of the challenges 
in the first year of running.) This produced a fully functional system which met all
the requirements of the experiment but also left many things implemented in a 
non-optimal way and made additional changes increasingly difficult to implement.

The new control system is described in detail in ~\cite{newCS-ref}.
Its tasks are clearly the same as for the old control system,
but it uses a distributed architecture and is built in a fully modular and 
extensible way. This modular framework allows for easy addition or 
reconfiguration of the CS, the necessity for which 
was found to be quite common in an active experiment. 

\section{Calibration and Reconstruction: A Two Pass System} \label{TwoPass}
During the first three years of running the calibration 
and reconstruction were done at the same time in a one pass system.
This placed constraints on the system which were not scalable and produced 
non-optimally calibrated data. For the fourth
year of running it was decided to change to a more classical two pass calibration 
and reconstruction system.

\subsection{Old One Pass System}
Through the end of the last run period, July 2002, the reconstruction and calibration 
were performed in a single processing pass of the data.

The runs had to be processed in the order they were taken since the `Rolling Calibrations' 
(RC), which track changes in the detector conditions with time, were calculated during 
the processing of the run. Most detector conditions change slowly and therefore information 
over a few runs (one run corresponds to about one hour of data taking) were combined to obtain 
enough statistics to make a good calculation. Reconstructed events were written to the 
event-store database continuously during processing, while the RC were written to the 
conditions database at the end of the run.

As \babar\ steadily increased its delivered luminosity and the reconstruction software 
increased in complexity, the processing time for a run became comparable or greater to the 
time taken to collect the data. The size of the farms (number of clients) could not scale 
to keep up with the incoming data. This was partly due to constraints from the event-store 
but also because managing a large number of clients ($>200$) posed its own problems. It became 
clear that this model would not scale sufficiently for the lifetime of the experiment.

Another disadvantage of the one pass system was that that the RC calculated during the 
reconstruction of run N were used as input to the reconstruction of run N+1, not run N. This 
did not provide the optimal calibration for detector quantities that can change quickly.

An advantage of this system was that the raw data only needed to be processed once. This avoided 
overhead (CPU and I/O) from multiple passes.

\subsection{New Two Pass System}
\begin{figure*}[t]
\centering
\includegraphics[width=135mm]{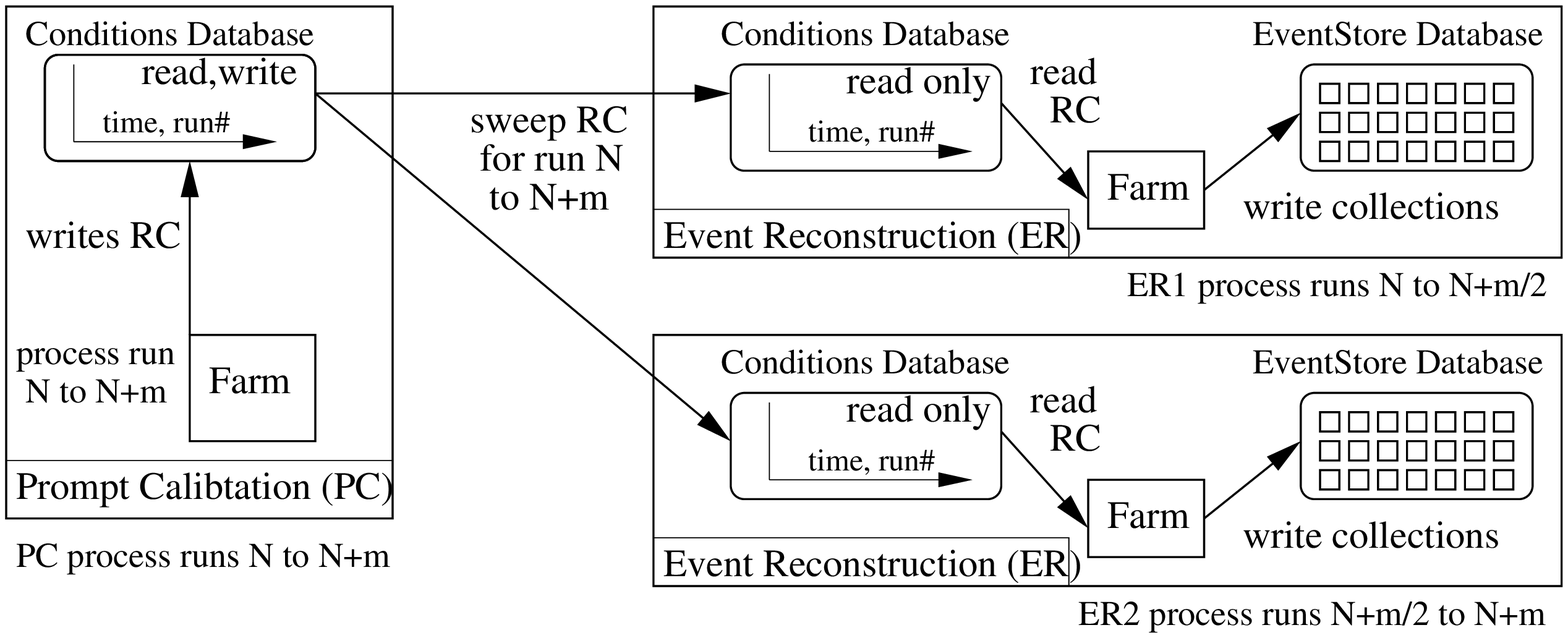}
\caption{Two Pass Multi-Farm System} \label{two_pass}
\end{figure*}

For the fourth year of data taking \babar\ has adopted a two pass calibration and 
reconstruction system, Figure~\ref{two_pass}.

The first pass, `Prompt Calibration' (PC), processes only a fraction ($1Hz$ fixed rate) of events 
in a run and writes out only the resulting conditions. It does not write to the event store.
The second pass, `Event Reconstruction' (ER), processes all selected events in a run, reading the 
conditions written by the PC pass, and writes the reconstructed events to the event-store.

The PC pass must process all runs in the order they were taken, as in the old one pass system, 
since it generates `Rolling Calibrations' (RC). The RC are written to the conditions federation 
at the end of processing run N, with a validity period starting at the beginning of run N.

The RC are then transfered from the PC conditions database to the ER conditions database. Runs 
can be processed in any order in the ER farm as long as the corresponding RC are present for that 
run. In this way run N is processed with the RC derived from run N, which provides the best 
reconstruction.

Since, in the PC pass, not all the events are processed, only enough to do the calibrations, the 
processing time is much less than the time needed to collect the data. In the current 
configuration, using 32 nodes, we are able to process the runs in about a half the time it 
takes to collect the data ($600 pb^{-1}$ vs. $300 pb^{-1}$ a day). The number of events needed 
for the calibration will not scale with luminosity, 
and therefore as the instantaneous luminosity increases the processing time for the runs 
will stay almost constant.

Since the  ER pass does not need to process the runs in order, runs calibrated in the same PC 
farm can be processed by one or more ER farms. All that is needed is that the RC be transfered 
to the appropriate farm(s).

This system is much more scalable than the one pass system. It combines a quick first pass which 
also provides prompt monitoring of the data quality with a full reconstruction pass that can be 
shared between more than one farm. See Section~\ref{numbers} for more details on the performance.

\section{Remote Reprocessing in Padova}
In 2001 the Padova INFN site agreed to build a data reprocessing center to help SLAC 
reprocess all the data collected in Run1, Run2, Run3 (and future runs). To realize such 
a reprocessing center all the software running at SLAC has been ported to a different platform 
(from Solaris to Linux) and site (different architecture of the farms, i.e. NFS usage is strongly reduced). 
The most important issues are discussed below, although not many technical details are described.

\subsection{The Control System, Site Dependence}
When Padova started running the reprocessing in September 2002 the new control system
mentioned above was still being tested at SLAC and so it was decided not to port it
to Padova right away, but rather run with the old control system in Padova.

The old control system scripts were strongly site-dependent. Much work was done to remove 
hardwired paths for directories and configurations and to make data storage directories 
also site-independent. The bookkeeping (relational) database SQL queries also needed to
be ported since SLAC uses Oracle and Padova uses MySQL.  
The two engines have many differences in SQL statements and native functions, especially 
when it comes to manipulating dates and times.

The data retrieval script which interfaces to the tape library had to be 
basically rewritten since the tape system in Padova was quite different than
the one at SLAC (HPSS).

\subsection{Solaris to Linux}
All of the servers at SLAC are Solaris (Sparc) machines while Padova
has only Linux (Intel) machines to capitalize on the ready availability of low price 
but high speed machines. This meant that all code to run the reconstruction
and communicate with the Objectivity/DB databases had to be ported to Linux.

This brought to light many problems, such as 
some c++ components which had endianess problems (Intel has a little endian for 
binary data, Sparc has a big endian). This affected the Logging Manager and Elf.

Problems were also found related to different file system behavior between Linux ext3 and Solaris UFS, 
especially when the Objectivity/DB databases are created and pre-sized with internal empty space 
(technically a ``hole-file''). Large file support in the Linux system was also an issue.
Many system packages from the standard Red Hat distribution had to be recompiled in Padova 
to handle files larger that $2GB$. Also some \babar\ 
configuration software had to be modified to support large files under Linux.

Many tests also had to be made to understand Objectivity/DB performance under Linux 
and many problem had to be solved. Furthermore, a good Objectivity/DB administration 
know-how was acquired at the Padova site.

\subsection{XTC Import}
A system to import all the XTC files has been completely home-made.
Several multi-stream ftp-like tools were tested, and bbcp~\ref{bbcp-ref}
is now used in production by the import system.

As time went by, some updating was necessary to optimize import times,
and to match the rising cartridges administration demand. A web form has
been developed to provide information about the import status, cartridge
information and mapping between XTC file and cartridges.

Initially the network performance was slow and compressing/decompressing
operations made the import faster but that is not currently necessary.
Currently, at any one time, the system is staging in an XTC file at
SLAC, importing up to three XTC files, in parallel, into a local buffer
area in Padova with a transfer rate of about 100 Mbits/sec,
and archiving one XTC file into the tape-library
in Padova.

Each independent part of the system runs continuously. When an error
happens, it retries at periodic time intervals, and when it still can not
succeed it finally sends a notification via e-mail. The XTC setup and recovery
after problems is fully automated and requires as little human intervention as
possible.

\subsection{Farm Monitoring}
Padova has developed a custom monitoring system for the computing farms. It runs in the background 
on a dedicated machine, and uses the Simple Network Management Protocol (SNMP) to query the status 
of each device on the farm. System Administrators must provide the monitor with an
XML configuration file containing the list of devices to monitor. 

\begin{figure}
\centering
\includegraphics[width=45mm,angle=-90]{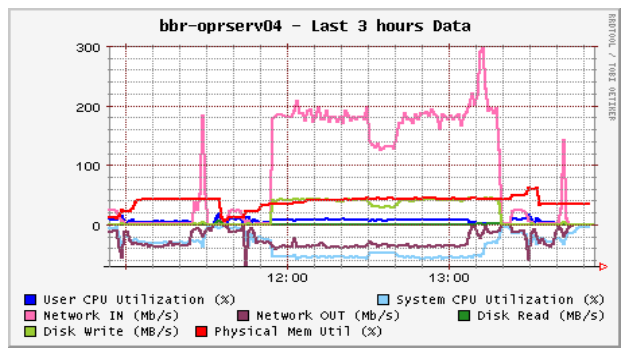}
\caption{Monitoring graph for a server in Padova.} \label{monitor}
\end{figure}

The monitoring system dynamically 
produces XML documents listing the status of each device; moreover, it can also create graphs 
showing the value of different monitored quantities as a function of time. Users can display status 
information by connecting to the web server which is embedded into the monitor. The quantities 
monitored are: cpu temperature, user and system cpu usage, memory usage, disk I/O and network I/O.
Figure~\ref{monitor} shows the monitoring graph for one of the main server machines in 
Padova.

\subsection{Data Export}
A system for exporting reprocessed data in Objectivity/DB format
to SLAC (and possibly other sites) has been developed in Padova.
This consists of a set of Perl scripts implementing a finite
state machine.

At the end of each production cycle (one week) production is
stopped, all databases are closed, then copied/attached to a QA
federation where they are checked for corruption, and then
eventually transferred to the remote site (SLAC). At every step a
relational database (MySQL) is updated with the status of each
database file.

\section{Current Processing}\label{numbers}

\subsection{Farm Configuration and Rates}
The configuration used for \babar\ Run3 processing and reprocessing is
as follows:

\begin{itemize}
\item 3 PC farms at SLAC 
  \begin{itemize}
    \item about 32 1.4GHz (Pentium III) cpus each
    \item 1 for the new data
    \item 2 for reprocessing 
  \end{itemize}
\item 5 ER farms at SLAC
  \begin{itemize}
    \item about 64 1.4GHz (Pentium III) cpus each
    \item 2 for the new data
    \item 3 for reprocessing 
  \end{itemize}
\item 4 ER farms in Padova
  \begin{itemize}
    \item about 80 1.26GHz (Pentium III) cpus each
    \item all 4 for reprocessing 
  \end{itemize}
\end{itemize}

As discussed earlier in this paper, one PC farm can provide the 
calibrations for multiple ER farms. Currently one PC farm feeds
two ER farms for the new data processing; one PC farm feeds three
ER farms for the reprocessing at SLAC; and one PC farm feeds
four ER farms for the reprocessing in Padova.

This is a distributed system. Not only are we transferring conditions
locally at SLAC between the PC and ER farms, but also between a PC 
farm at SLAC and the ER farms in Padova.

In the current setup each PC farm can process up to $600 pb^{-1}$
per day while each ER farm can do about $150 pb^{-1}$. Both types
of farms have significant deadtime between runs. For the PC farms 
this deadtime comes from the calculation of the RC at the end 
of the run and the writing to the conditions database. For the 
ER farms the deadtime is bigger and mainly due to overhead from 
setting up and then cleaning up the Objectivity/DB event-store 
database. To optimize the cpu usage, Monte Carlo (MC) production jobs
are run in the background with a large nice value. Therefore during
the down time between runs and any other longer down time the MC 
jobs take over the cpus, while during processing they do not impact
the Elf processing jobs.  

\begin{figure}
\centering
\includegraphics[width=70mm]{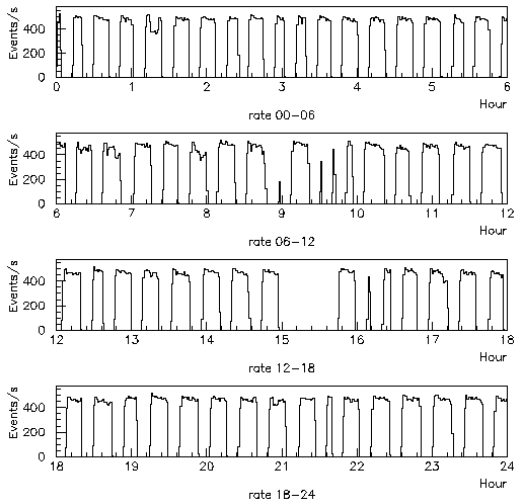}
\caption{The input event processing rate in a PC farm over 24 hours.} \label{pcRate}
\end{figure}

\begin{figure}
\centering
\includegraphics[width=70mm]{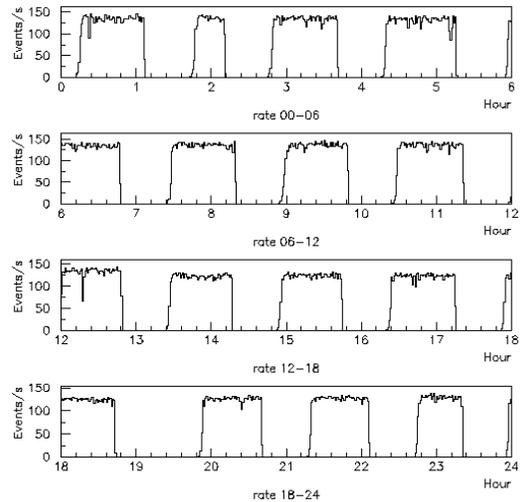}
\caption{The input event processing rate in an ER farm over 24 hours.} \label{erRate}
\end{figure}

This deadtime is not all irreducible and therefore allows room for 
improvement of the production rate. For the newly collected data
we expect the PC pass to be done within 8 to 10 hours and the
ER pass within 24 hours. Figures ~\ref{pcRate} and ~\ref{erRate}
show the PC and ER pass (input) processing rates over a 24 hour period.

\subsection{Outstanding Issues}
The control system and related activities are still being constantly 
improved. In this first year of running with the new control system
many new features were added as needed and further additions or
changes are anticipated to optimize the performance and increase the
ease of operation. The goal is to have a stable system which needs
as little human intervention as possible.

There is a constant struggle to make sure that the appropriate 
monitoring is available to spot problems with reconstruction and calibrations.
This often requires additions to the validation aspect of the
control system when a new problem is found, but input from detector 
subsystems on the data quality is very important.

We have had a number of problems with data corruption, which meant
that a subset of the reprocessing had to be redone. Some corruption
was due to hardware problems and some due to Objectivity/DB implementation.

Although what was mainly discussed in this paper was the data 
processing or reprocessing, there is another step before the data
is available to the physics analysis users. The data written to
the dedicated production servers is migrated to the mass storage
system and also must be ``swept'' to dedicated analysis servers.
Due to current limitations of the Bdb/Objectivity/DB event-store
implementation, data is not available for up to 10 days after it is
processed.

\section{Summary and Outlook}
The \babar\ Prompt Reconstruction system is used for processing of new
data and reprocessing of the data set. In the current year of running
we have moved from a single pass architecture to a two pass, calibration
and reconstruction system. This provides better calibrated data and
allows for further scaling to keep up with increases in the total 
data set.

In order to exploit available resources, we have moved to a distributed
system using farms in multiple sites, SLAC and Padova. This allows
sharing not only of hardware but also of personnel resources.

Significant progress has been made in building a stable, scalable 
system. We now believe that we have an architecture better positioned
to scale well through future luminosity upgrades.




\end{document}